\PassOptionsToPackage{table,prologue}{xcolor}

\documentclass[sigconf, nonacm]{acmart}

\usepackage{tikz}
\usetikzlibrary{fit}
\usetikzlibrary{matrix}

\usepackage{algorithm}
\usepackage{algpseudocode}

\usepackage{graphicx}
\usepackage{subcaption}

\usepackage{balance}
\usepackage{flushend}

\newcommand{\dataset}[1]{\textsf{#1}} % Usa small caps
\newcommand{\compressor}[1]{\textsc{#1}} % Usa sans-serif
\newcommand{\algorithmname}[1]{OnPair#1}
\newcommand{\substring}[1]{\texttt{``#1''}}



\newcommand\vldbpagestyle{plain} 

\begin{document}
\title{\algorithmname{}: Short Strings Compression for Fast Random Access}
%%
%% The "author" command and its associated commands are used to define the authors and their affiliations.
\author{Francesco Gargiulo}
\affiliation{%
  \institution{University of Pisa}
  \city{Pisa}
  \state{Italy}
}
\email{francesco.gargiulo@phd.unipi.it}

\author{Rossano Venturini}
\affiliation{%
  \institution{University of Pisa}
  \city{Pisa}
  \state{Italy}
}
\email{rossano.venturini@unipi.it}

\begin{abstract}
We present \compressor{\algorithmname{}}, a dictionary-based compression algorithm designed to meet the needs of in-memory database systems that require both high compression and fast random access. Existing methods either achieve strong compression ratios at significant computational and memory cost (e.g., \compressor{BPE}) or prioritize speed at the expense of compression quality (e.g., \compressor{FSST}). \compressor{\algorithmname{}} bridges this gap by employing a cache-friendly dictionary construction technique that incrementally merges frequent adjacent substrings in a single sequential pass over a data sample. This enables fast, memory-efficient training without tracking global pair positions, as required by traditional \compressor{BPE}. We also introduce \compressor{\algorithmname{16}}, a variant that limits dictionary entries to 16 bytes, enabling faster parsing via optimized longest prefix matching. Both variants compress strings independently, supporting fine-grained random access without block-level overhead. Experiments on real-world datasets show that \compressor{\algorithmname{}} and \compressor{\algorithmname{16}} achieve compression ratios comparable to \compressor{BPE} while significantly improving compression speed and memory usage.
%We present \compressor{\algorithmname{}}, a dictionary-based compression algorithm designed for datasets of strings requiring efficient random access. Existing methods either offer high compression ratios at the cost of compression speed and memory usage (e.g., \compressor{BPE}), or prioritize fast performance with limited compression effectiveness (e.g., \compressor{FSST}). \compressor{\algorithmname{}} bridges this gap with a cache-aware, memory-efficient approach to dictionary construction. It generates new dictionary entries by pairing frequently co-occurring substrings during a single sequential pass over a data sample, avoiding the need to maintain and update global pair counts and positions as in traditional \compressor{BPE}. Moreover, fast decompression is achieved through SIMD-friendly decoding. We also introduce \compressor{\algorithmname{16}}, a variant that limits dictionary entries to 16 bytes, enabling faster parsing via optimized prefix matching. Both variants compress strings independently, supporting fine-grained random access without block-level overhead. Experiments on real-world datasets show that \compressor{\algorithmname{}} and \compressor{\algorithmname{16}} achieve compression ratios comparable to \compressor{BPE} while significantly improving compression speed and memory usage. 
\end{abstract}

\maketitle

\pagestyle{\vldbpagestyle}

\section{Introduction}
Main memory database systems are essential in data-intensive applications where low-latency access and high throughput are crucial (see \cite{faerber2017mainmemory} and references therein). Systems like SAP HANA \cite{faerber2012sap}, HyPer \cite{kemper2011hyper}, and MonetDB \cite{idreos2012monetdb} eliminate traditional disk I/O bottlenecks by retaining data in DRAM, significantly accelerating query processing. However, a critical bottleneck remains: the rapid growth in data volumes continues to outpace improvements in DRAM capacity and cost-efficiency. As a result, storing large datasets entirely in memory without incurring significant hardware costs becomes increasingly infeasible.

Data compression emerges as a natural remedy to this problem, with the dual goals of reducing memory footprint and preserving—or ideally accelerating—query performance. In database systems, compression techniques can be applied at varying granularities, ranging from fine-grained approaches (individual fields) to coarser-grained methods (records, columns, or entire blocks). However, not all compression strategies are equally suited for in-memory workloads, especially when fast random access to individual data entries is needed. Compression schemes can be broadly classified into two types:
\begin{itemize}
    \item Lightweight compression algorithms \cite{abadi2006compression, damme2017lightweight} are typically designed to prioritize decompression speed and direct operability on compressed data, trading off higher compression ratios for faster execution. Examples include \textit{frame-of-reference}, \textit{delta encoding}, \textit{domain encoding}, \textit{run-length encoding}, and \textit{null suppression}. These methods are particularly effective in columnar storage layouts, where data locality allows for more aggressive optimizations. 
    \item Heavyweight compression algorithms, such as LZ77-based methods (like \compressor{gzip} \cite{deutsch1996gzip} and \compressor{Zstandard} \cite{collet2018zstandard}), operate at coarser levels. These achieve higher compression ratios by leveraging redundancy across larger data segments. However, they are oblivious to the structure of the underlying data, meaning that it is not possible to interpret (and thus operate on) the compressed bytes.
\end{itemize}
For string data specifically, domain encoding is widely used: it replaces variable-length strings with fixed-size identifiers that reference a dictionary. If the number of unique strings is small, this may yield substantial space savings. Compression can be further improved by using compressed dictionaries, and order-preserving dictionaries even support the direct execution of range queries without decompressing the actual data \cite{liu2019dictionaries, binnig2009dictionary}.

While domain encoding is highly effective, it is only applicable when the number of distinct values in a column is relatively small. In many practical scenarios, however, string columns contain a large number of unique entries, each of which may be short and infrequently repeated. In such cases, conventional domain encoding becomes ineffective. Similarly, general-purpose compression algorithms based on LZ77 are often unsuitable, as they require larger input sizes to discover redundancy and achieve meaningful compression. These approaches are not optimized for handling numerous small strings independently, which is common in real-world workloads.

To address these limitations, \textit{Fast Static Symbol Table} (\compressor{FSST}) \cite{boncz2020fsst} was introduced as a field-level string compression scheme designed for high-speed encoding and efficient random access. \compressor{FSST} employs an external dictionary to encode each string independently, substituting frequent substrings (up to 8 bytes in length) with compact single-byte tokens. However, \compressor{FSST}'s dictionary construction process is heavily biased toward speed, which limits its ability to discover and exploit longer or more complex patterns in the data—ultimately resulting in lower compression levels.

On the other hand, \textit{Byte-Pair Encoding} (\compressor{BPE}) \cite{gage1994bpe} achieves high compression ratios by iteratively merging the most frequent adjacent symbol pairs in the dataset. This results in dictionaries that effectively capture long and recurring patterns, making \compressor{BPE} highly effective at compressing string data. Like \compressor{FSST}, \compressor{BPE} is also naturally suited for scenarios that require fast random access, as each string can be compressed independently. However, despite its compression effectiveness, \compressor{BPE} is often impractical for large-scale string compression due to its substantial computational and memory overhead. Building the dictionary requires repeatedly scanning and updating global frequency statistics, which involves frequent, non-local accesses to token positions across the dataset—an operation that is not cache-friendly. In addition, the algorithm must explicitly track the positions of all token pairs in order to apply replacements efficiently, which incurs significant memory usage. 

In this paper, we present \compressor{\algorithmname{}} and its optimized variant \compressor{\algorithmname{16}}, two compression algorithms specifically designed to combine strong compression ratios with fast, fine-grained random access. The main innovations of this work are:
\begin{itemize}
    \item \textit{A scalable, cache-friendly dictionary construction algorithm}. Unlike traditional \compressor{BPE}, which relies on costly (both in time and memory) global frequency statistics, \compressor{\algorithmname{}} incrementally merges frequent adjacent token pairs during a single sequential pass over a data sample, enabling a fast and memory-efficient dictionary population.
    \item \textit{An optimized longest prefix matching algorithm}. The fixed token length in \compressor{\algorithmname{16}} enables several implementation-level optimizations—including bitwise comparisons, inlined suffixes, and efficient bucket indexing—that improve compression performance.
    \item \textit{A cache-aware design for high decompression speed}. The dictionary is compact enough to fit in the L2 cache and decoding is a predictable, SIMD-friendly process.
\end{itemize}
Our experiments show that \compressor{\algorithmname{}} and \compressor{\algorithmname{16}} match the compression effectiveness of \compressor{BPE} while being significantly faster and more memory-efficient to train. This makes them especially well-suited for large-scale, in-memory workloads where traditional \compressor{BPE} becomes impractical. \compressor{\algorithmname{16}}, in particular, offers an attractive speed-quality trade-off: it retains nearly the same compression ratio as \compressor{\algorithmname{}}, but benefits from hardware-friendly design choices that accelerate both compression and decompression.

\section{Related Work}
In this section, we discuss the primary approaches to the problem, distinguishing between block-based solutions that group strings together and methods that allow independent decompression of strings. Each approach is analyzed in terms of its benefits and drawbacks, particularly for use cases requiring fast random access to individual strings.

\subsection{Block-Based Solutions}
General-purpose compressors like \compressor{gzip} \cite{deutsch1996gzip}, \compressor{LZ4} \cite{collet2011lz4}, and \compressor{Zstandard} \cite{collet2018zstandard} combine ideas from LZ77 compression \cite{ziv1977universal} with entropy coding techniques, typically utilizing variants of \textit{Huffman coding} \cite{huffman1952codes} or \textit{Asymmetric Numeral Systems} \cite{duda2013ans}. 

LZ77 is a dictionary-based compression algorithm that exploits repeated sequences of data by replacing them with references to previous occurrences within a sliding window. This approach is highly effective for data with repetitive patterns, as it replaces redundant sequences with compact back-references.

However, this compression efficiency comes at a cost. One major drawback of LZ77-based methods is their lack of native support for fast random access. Later parts of the compressed stream may refer back to earlier parts which may be themselves references to earlier ones. This means that decompressing a particular segment often requires recursively resolving a chain of back-references. This chained dependency limits the ability to jump directly to and extract a specific portion of the original data without decoding a large preceding context. Moreover, LZ77 performs poorly on small strings or small datasets, because the opportunity to find repeated patterns within a limited sliding window is greatly reduced. With fewer repeated sequences to exploit, the algorithm often emits many short literals instead of back-references, leading to minimal compression gains—or even overhead due to metadata.

To address these limitations, a common strategy involves grouping short strings into fixed-size blocks. This class of solutions, which we refer to as \textit{Block-Based Compressors}, relies on processing data in blocks of a maximum predefined size. By aggregating many short strings into a single block, LZ77 and similar algorithms can discover and exploit cross-string redundancy, improving compression ratios.

A key drawback of block-based compression is that accessing a single element potentially requires decompressing the entire block. This is particularly problematic when frequent random access to individual strings is required. There is an inherent trade-off between compression ratio and random access speed when selecting the block size: smaller blocks enable faster access but yield lower compression ratios, while larger blocks offer better compression but slow down selective decompression. Consequently, block-based methods struggle to efficiently serve use cases that demand both high compression and fast, fine-grained access to individual strings.

\subsection{Byte-Pair Encoding}
\textit{Byte-Pair Encoding} (\compressor{BPE}) \cite{gage1994bpe} is a data compression algorithm that iteratively identifies the most frequent pair of adjacent bytes in the data and replaces all its occurrences with a new byte value that does not appear in the original data. This substitution process is repeated until no pairs remain or all available byte values have been utilized. To decompress the data, a lookup table maps each substituted byte back to its original pair. The major limitation of this approach is the restricted number of non-used byte values available for substitutions, which constrains the algorithm's practical applicability. Nonetheless, this idea of iteratively substituting the most frequent pair of adjacent symbols is used in many applications.

A more recent adaptation of the algorithm \cite{sennrich2016subword} has been widely used for subword tokenization in natural language processing tasks. Instead of merging frequent pairs of bytes, this variant merges pairs of characters. The process begins with a dictionary initialized to the character vocabulary. The algorithm then iteratively identifies the most frequent adjacent symbol pairs (where symbols may initially be characters) and adds them as new entries to the dictionary. This iterative merging continues until a predefined maximum dictionary size is reached. % A key feature of this approach is its restriction to within-word boundaries, which ensures that frequent subword units are identified without merging symbols across different words. This property makes the method particularly effective for natural language processing applications, as the resulting dictionary captures meaningful subword units, such as prefixes, suffixes, and stems. These subword units enable the model to handle out-of-vocabulary words by representing them as compositions of known subwords.

In practice, \compressor{BPE} effectively identifies meaningful subsequences and achieves competitive compression ratios, with theoretical analyses of the algorithm provided in \cite{zouhar2024formalbpe, kozma2024theoreticalbpe}. However, the algorithm has significant limitations. Introducing a new symbol requires substantial computational effort to update the frequencies of all affected pairs, a process that becomes increasingly costly as the dataset size grows. Additionally, the frequent pairs may be scattered throughout the input, resulting in poor cache utilization during these updates. Maintaining a complete record of pair positions also demands considerable memory, making the algorithm impractical for large datasets.

\subsection{Re-Pair}
\compressor{Re-Pair} \cite{larsson1999repair} is a grammar-based compression algorithm closely related to \compressor{BPE}. \compressor{Re-Pair} iteratively identifies the most frequent pair of adjacent symbols, say $ab$, and replaces it with a new symbol, say $X$, adding the production rule $X\rightarrow ab$ to the grammar. Unlike \compressor{BPE}, which halts once a predefined dictionary size is reached, \compressor{Re-Pair} continues this process until no pair of symbols appears more than once. The resulting grammar becomes an integral part of the compressed data and must itself be efficiently encoded to reduce storage overhead. Substantial research has focused on improving and analyzing \compressor{Re-Pair}. This includes theoretical studies on its compression efficiency \cite{navarro2008repair} and efforts to lower its memory requirements \cite{bille2016spaceefficient, gagie2019rpair}. Notably, grammar-based compression techniques, including \compressor{Re-Pair}, enable direct access to substrings in logarithmic time relative to the size of the original data \cite{bille2010random}. While \compressor{Re-Pair} offers strong compression performance, it inherits significant computational and memory overhead challenges from \compressor{BPE}. Moreover, accessing an item requires recursively expanding the corresponding grammar rules, which is less straightforward and typically slower in practice than the direct dictionary lookups used in \compressor{BPE}.

\subsection{Fast Static Symbol Table}
\textit{Fast Static Symbol Table} (\compressor{FSST}) \cite{boncz2020fsst} is a lightweight compression schema designed specifically for string data, offering random access to individual compressed strings. \compressor{FSST} achieves this by replacing frequently occurring substrings (up to 8 bytes in length) with 1-byte codes, using a dictionary that maps these codes to their corresponding substrings. Limiting the maximum length of strings in the dictionary to 8 bytes has a twofold advantage: it allows the dictionary to remain in the L1 cache and enables the use of unaligned store operations to quickly copy each substring in a single operation during decompression. %The dictionary reserves a symbol as an escape code to handle single bytes for which there is no corresponding symbol. 

To generate the dictionary, \compressor{FSST} uses a bottom-up approach that refines the dictionary over multiple iterations. Each iteration consists of two parts:
\begin{enumerate}
    \item The input data is tokenized using the current dictionary, selecting the longest possible matches.
    \item A new dictionary is generated by selecting the most frequent symbols based on their contribution to compression, computed as frequency multiplied by substring length. Candidate symbols include those from the previous iteration and new ones formed by merging frequently occurring pairs.
\end{enumerate}
To improve efficiency, \compressor{FSST} samples the input data during dictionary construction and applies lossy perfect hashing to accelerate symbol lookups. Additionally, \compressor{FSST} leverages AVX-512 SIMD instructions to enhance encoding speed.

Overall, \compressor{FSST} offers very fast compression and decompression speeds while enabling random access to individual strings. However, its compression ratio is relatively low—a direct consequence of its design, which prioritizes speed over compression effectiveness. While this trade-off is acceptable in latency-critical applications, many real-world scenarios follow a compress-once, decompress-many pattern. In such cases, it is often worthwhile to sacrifice some compression speed to achieve better compression ratios.

\section{Proposed Algorithm}
\label{sec:proposed_algorithm}

\subsection{Overview}
\compressor{\algorithmname{}} is a dictionary-based compression algorithm designed for string datasets that require efficient random access. The compression process consists of two distinct phases:
\begin{itemize}
    \item \textit{Training Phase}: The algorithm builds the dictionary by scanning the data sequentially, using a longest prefix matching strategy to parse the input and identify frequent adjacent token pairs. When the frequency of a pair exceeds a predefined threshold, a new token is created to represent the merged pair. This continues until the dictionary is full or the input data is exhausted. The dictionary supports up to 65{,}536 tokens, with each token assigned a fixed 2-byte ID. 
    \item \textit{Parsing Phase}: Once the dictionary is constructed, each string is compressed independently into a sequence of token IDs by greedily applying longest prefix matching.
\end{itemize}
\compressor{\algorithmname{16}} is a variant that limits dictionary entries to a maximum length of 16 bytes. This constraint enables further optimizations in both longest prefix matching and decoding.

Throughout this section, we illustrate key design choices and their impact using experimental results from the \dataset{Book Titles} dataset. This dataset reflects a realistic database workload: it consists of moderately short strings, contains meaningful redundancy, and is representative of catalog-style data commonly stored in main-memory systems. See \autoref{sec:experiments} for full details on datasets and experimental setup.

\subsection{Training Phase}
During the training phase, \compressor{\algorithmname{}} constructs a dictionary of frequent substrings through an iterative process. To avoid overfitting to the beginning of the dataset, the sample of data processed during training is a randomly selected subset of strings, shuffled to ensure a random processing order. This exposes the algorithm to a more representative distribution of patterns across the full dataset, ensuring that the dictionary reflects global, rather than local, properties of the data. This is particularly important since the dictionary construction halts once the maximum size is reached, potentially biasing the dictionary toward early patterns if the processing order is not randomized. 

The training process continues until either the maximum dictionary size is reached (65{,}536 tokens), or if there is no more data to process:
\begin{enumerate}
    \item The dictionary is initially populated with 256 entries, one for each possible byte value (0–255), with each byte directly serving as its own token ID.
    \item The algorithm scans through the data, one string at a time, using a longest prefix matcher to identify existing matches.
    \item When pairs of tokens appear together frequently (reaching a predefined threshold), they are merged into a new token and added to the dictionary. After the merge, the last parsed token is replaced with the newly created token, ensuring that subsequent pair counting continues with the new token rather than the second token of the original pair.
\end{enumerate}

\begin{figure}
    \centering
    \begin{tikzpicture}[
    box/.style={draw=gray!50, minimum height=0.7cm, minimum width=0.7cm, 
                anchor=south west, fill=white, line width=0.5pt},
    token/.style={draw=gray!50, rounded corners=3pt, minimum height=0.6cm, 
                 fill=blue!10, font=\small\ttfamily, align=center, text width=2.5cm},
    arrow/.style={->, >=stealth, thick, draw=gray!70},
    freq/.style={draw=gray!50, rounded corners=3pt, fill=orange!10, 
                minimum width=0.8cm, minimum height=0.6cm,
                text depth=0.25ex, text height=1.5ex, font=\small\ttfamily},
    threshold/.style={draw=gray!50, rounded corners=3pt, fill=orange!10, 
                     minimum width=2.5cm, 
                     text depth=0.25ex, text height=1.5ex},
    dict/.style={matrix of nodes, nodes={rectangle, draw=gray!30, 
                text width=2.5cm, align=center, anchor=center},
                row sep=0.2cm, column sep=0.2cm, draw=gray!50, rounded corners=3pt},
    dict_header/.style={fill=blue!10},
    dict_value/.style={fill=orange!10},
    highlight/.style={draw=#1, thick, rounded corners=3pt, fill=#1!10, opacity=0.4},
    section/.style={font=\normalsize\bfseries, align=center, text width=4cm},
    annotation/.style={font=\small\itshape, text=gray!80}
]
    
    % Define fixed positions for better spacing
    \def\centerhorizontal{4.0}
    
    % SECTION 1: Longest Prefix Matching
    \node[section, anchor=center] at (\centerhorizontal,0.1) {Longest Prefix Match};
    
    % Input string with vertical separators 
    \foreach \x/\c [count=\i from 0] in {0.5/a, 1.2/b, 1.9/r, 2.6/a, 3.3/c, 4.0/a, 4.7/d, 5.4/a, 6.1/b, 6.8/r, 7.5/a} {
        \node[box] at (\x,-1) {\c};
        \draw[gray!30] (\x+0.7,-1) -- (\x+0.7,-0.3);
    }
    
    % Previous match highlight
    \node[highlight=blue, fit={(0.5,-1.0) (3.23,-0.30)}, inner sep=2pt] (prev_match) {};
    \node[annotation] at (1.55,-1.5) {Previous match};
    
    % Current match highlight
    \node[highlight=red, fit={(3.37,-1.0) (5.34,-0.3)}, inner sep=2pt] (curr_match) {};
    \node[annotation] at (4.335,-1.5) {Current match};
    
    % Token IDs
    \node[token] (token1) at (1.55,-2.6) {Token ID: 512};
    \node[token] (token2) at (4.5,-2.6) {Token ID: 278};
    
    % Draw arrows connecting matches to tokens
    \draw[arrow] (prev_match) -- (token1.north);
    \draw[arrow] (curr_match) -- (token2.north);
    
    % SECTION 2: Frequency Update
    \node[section, anchor=center] at (\centerhorizontal,-3.5) {Frequency Update};
    
    % Frequency tracking
    \node[freq] (freq_label) at (1.5,-4.13) {Pair \texttt{(512,278)}:};
    \node[freq] (freq_old) at (3.5,-4.13) {9};
    \node[freq] (freq_new) at (5,-4.13) {10};
    \draw[arrow] (freq_old) -- node[annotation, above, font=\small, pos=0.45] {+1} (freq_new);
    
    % Threshold check
    \node[threshold] (threshold) at (1.5,-4.8) {Threshold: 10};
    \node[annotation] at (5.1,-4.82) {Frequency reached, create new token};
    
    % SECTION 3: Dictionary Insertion (add more space below threshold)
    \node[section, anchor=center] at (\centerhorizontal,-5.65) {Dictionary Insertion};
    
    % Dictionary table
    \matrix (dict_table) [matrix of nodes, 
        nodes={draw=gray!50, rounded corners=3pt, fill=blue!3, 
               minimum width=1.5cm, minimum height=0.5cm, align=center, 
               font=\footnotesize\ttfamily, text depth=0.1ex, text height=1.5ex},
        row sep=0.1cm, column sep=0.3cm,
        inner sep=0.1cm, draw=gray!50, rounded corners=5pt,] at (\centerhorizontal,-7.55) {
        \textbf{Token ID} & \textbf{Entry} \\
        \dots & \dots \\
        278 & \substring{cad} \\
        512 & \substring{abra} \\
        793 & \substring{abracad} \\
    };

    \end{tikzpicture}
    \caption{Illustration of the dictionary construction process in \compressor{\algorithmname{}} during its training phase. Here, we assume that \substring{abra} (Token ID: 512) and \substring{cad} (Token ID: 278) are the last two longest matches in the input, for the current dictionary. When the frequency of a token pair (512, 278) reaches the threshold, a new token is created, representing their concatenation (\substring{abracad}). Then, the previous match is updated with this new token for subsequent pair counting.}
    \label{fig:dictionary_construction}
\end{figure}
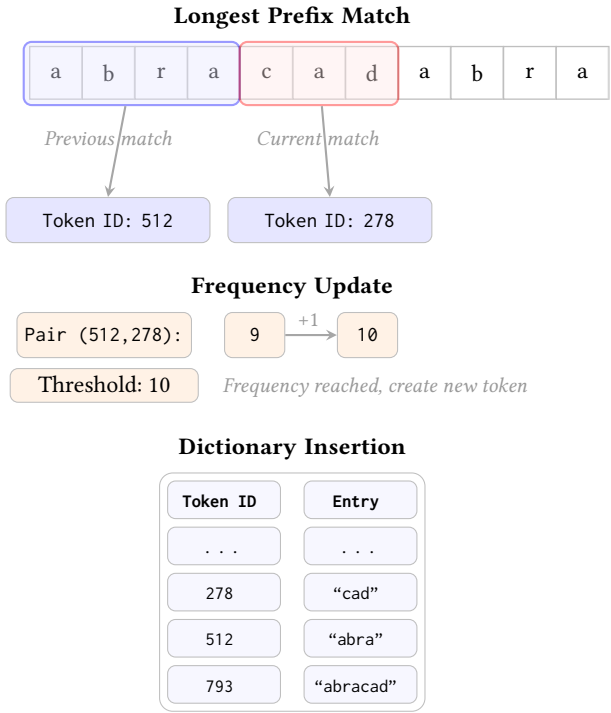

\autoref{fig:dictionary_construction} provides a visual representation of the dictionary construction process, showing how the frequency of adjacent token pairs is tracked and used to create new tokens when the threshold is reached.

\subsubsection{Pair Frequency Threshold}
The pair frequency threshold sets the minimum number of occurrences required to merge a token pair during dictionary construction. Higher thresholds slow dictionary growth but favor globally frequent patterns, potentially improving compression at the cost of more processing. Lower thresholds populate the dictionary quickly—useful for small datasets or latency-sensitive scenarios—but risk capturing unrepresentative patterns. If the threshold is too high, few pairs may qualify, leaving the dictionary underutilized.

\begin{figure}
  \centering
  \includegraphics[width=\linewidth]{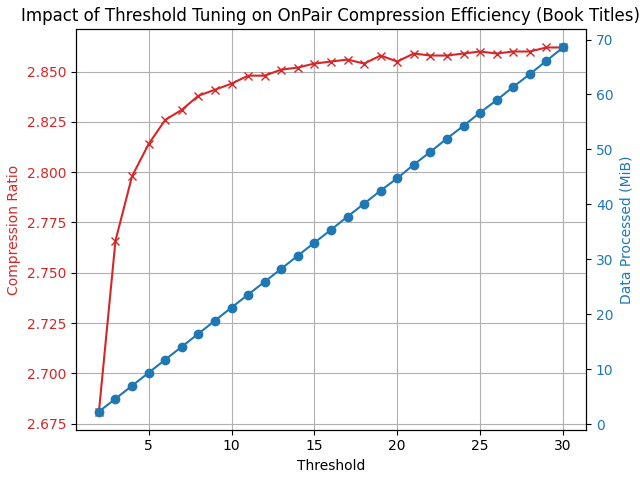}
  \caption{Compression ratio and training data volume as functions of the pair frequency threshold (2–30), measured during dictionary construction for \compressor{\algorithmname{}} on the \dataset{Book Titles} dataset.}
  \label{fig:threshold_experiment}
\end{figure}

\autoref{fig:threshold_experiment} shows how the threshold impacts both compression ratio and the amount of data required to fill the dictionary. Lower thresholds yield rapid initial improvements, but beyond a certain point, the compression benefit plateaus while training cost continues to rise, making higher thresholds less advantageous. 

The threshold is never set to 1, as this would cause the algorithm to add a growing sequence of prefixes for each string. Each newly merged token replaces the last matched token, so frequency tracking resumes with it—effectively generating a series of increasingly longer prefixes that are unlikely to repeat.

To adapt the threshold to the dataset size, we use a slowly growing function, such as a logarithm:
\[
\text{threshold} = \max \left( {2, \lfloor\log_2{(S)} \rfloor} \right)
\]
where $S$ is the dataset size in MiB. This ensures small datasets use low thresholds for fast gains, while larger ones adopt more selective merges. Thus, logarithmic growth allows us to balance compression quality and training cost.

\subsubsection{Limiting Dictionary Entry Size}
Limiting the maximum length of dictionary entries provides two key performance benefits:
\begin{itemize}
    \item \textit{Compression Speed}: Long entries can degrade parsing performance, especially when they are rarely used. This issue is particularly evident in datasets with many strings sharing long common prefixes that only diverge after several characters, such as groups of URLs from the same domain. Since the parser favors the longest possible match, it must repeatedly compare against these long entries, resulting in many wasted comparisons and significant slowdowns.
    \item \textit{Decompression Speed}: Modern processors support SIMD instructions that can load or store multiple bytes in a single operation. For example, AVX-512 can process up to 64 bytes at once. By limiting dictionary entries to a fixed maximum size, decompression becomes predictable: the decoder can copy each token using a single SIMD instruction and then advance the output buffer by the token's actual length.
\end{itemize}
\compressor{\algorithmname{16}} is a variant of the algorithm that restricts dictionary entries to a maximum of 16 bytes—large enough to capture most recurring patterns, while enabling key prefix-matching optimizations. In particular, strings longer than 8 bytes can be represented as two 64-bit integers, allowing for fast bitwise comparisons. Additionally, the first few longest suffixes can be inlined directly into the lookup structure, avoiding costly memory indirections. Details are provided in \autoref{sec:lpm}.

To quantify the contribution of each token to compression, we define a simple metric called \textit{token gain}. For a token $t$ with length $\ell(t)$ (in bytes) and frequency $f(t)$:
\[
token\_gain(t) = (\ell(t) - 2) \times f(t) - \ell(t)
\]
The first term reflects the saved space from replacing raw substrings with 2-byte token IDs, while the second accounts for the dictionary space required to store the token's content.

\begin{figure}
  \centering
  \includegraphics[width=\linewidth]{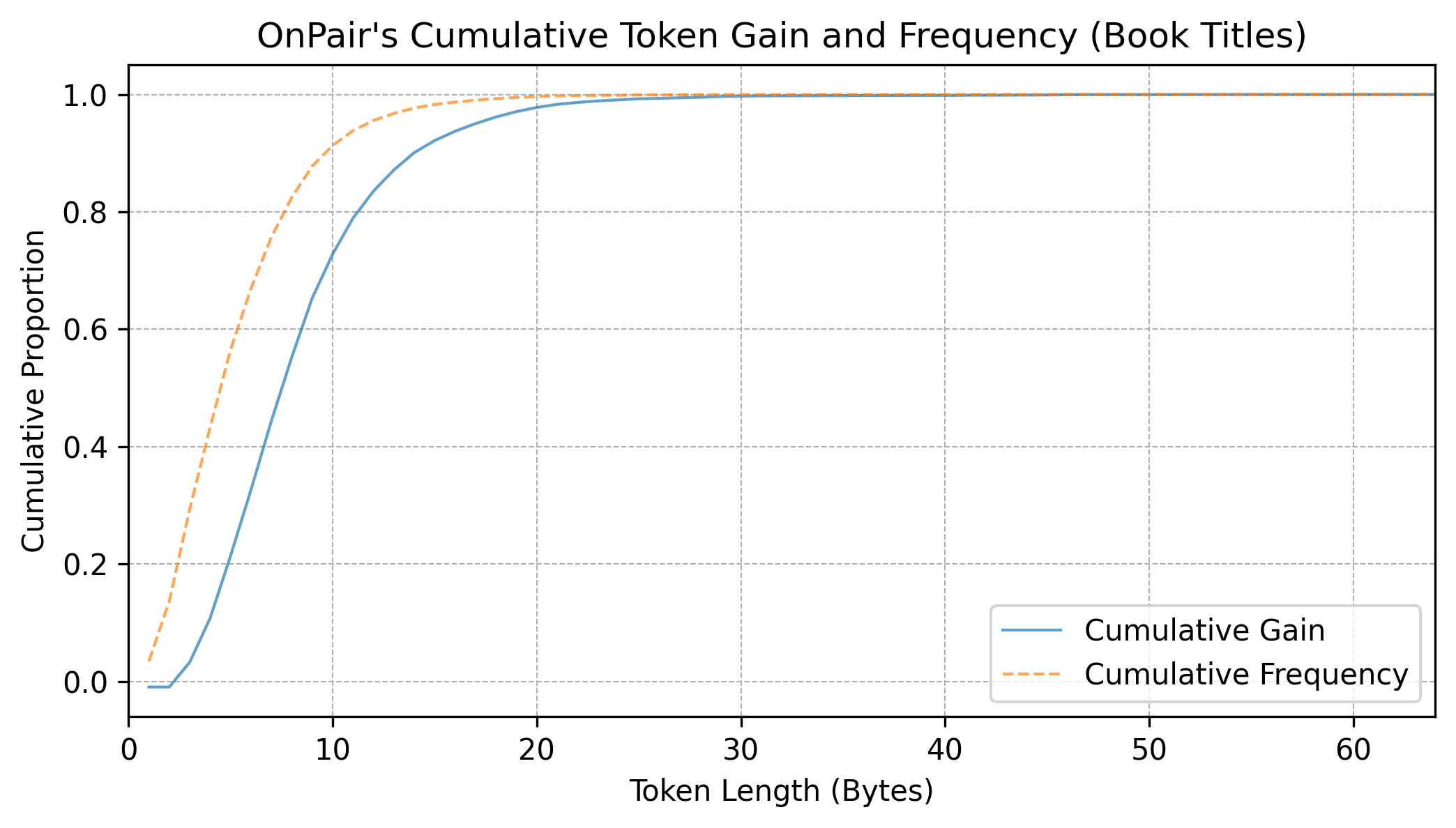}
  \caption{Cumulative gain and frequency by token length for \compressor{\algorithmname{}} on the \dataset{Book Titles} dataset. The majority of redundancy is captured by relatively short tokens: tokens up to 16 bytes account for over 93\% of the total gain and nearly 99\% of all token occurrences.}
  \label{fig:token_length_distribution_book_titles_onpair}
\end{figure}

As shown in \autoref{fig:token_length_distribution_book_titles_onpair}, on the \dataset{Book Titles} dataset, tokens up to 16 bytes account for over 93\% of the total gain, indicating that this limit captures the vast majority of useful patterns in practice—though the exact trade-off naturally depends on the dataset.

This 16-byte limit also provides a strict upper bound on dictionary memory footprint. In the worst case, where all $2^{16}$ entries reach the maximum length, the dictionary occupies at most $2^{16} \times 16 = 1$~MiB for the data section. Additionally, the offset array requires $2^{16}$ entries, each stored as a 4-byte integer, for a total of 0.25~MiB. This yields a combined upper bound of 1.25~MiB. However, as shown in \autoref{sec:dictionary_memory_footprint}, observed dictionary memory usage is typically smaller.

\subsection{Parsing Phase}
Once the dictionary is built, the parsing phase efficiently compresses the input data into a sequence of 2-byte token IDs that represent the original strings in compressed form. For each string in the dataset, the algorithm iteratively repeats the following operations:
\begin{enumerate}
    \item Find the longest possible dictionary match at the current position.
    \item Add the corresponding token ID to the compressed output.
    \item Advance the position by the length of the matched pattern.
\end{enumerate}
Since the dictionary is usually constructed from a small sample of the dataset, the parsing phase typically accounts for the majority of the overall compression time. The longest prefix matching algorithm is the central component of both training and parsing, and its efficiency is therefore the primary driver of overall compression performance.

\subsection{Longest Prefix Matching} \label{sec:lpm}
The \textit{longest prefix matching} (LPM) algorithm is a fundamental component of both \compressor{\algorithmname{}} and \compressor{\algorithmname{16}}, designed to efficiently identify the longest possible dictionary match for a given position in the input data. Since this is the most computationally intensive operation of the compression algorithm, the efficiency of the LPM directly impacts the overall performance of the compression process.

\subsubsection{Data Structures}
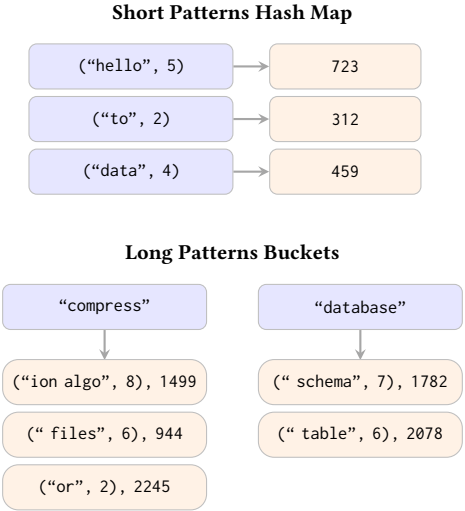
\begin{figure}
    \centering
    \begin{tikzpicture}[
        map_key/.style={draw=gray!50, fill=blue!10, rounded corners=3pt, minimum height=0.6cm, 
                        font=\ttfamily\footnotesize, text width=2.5cm, align=center},  % Reduced width
        map_value/.style={draw=gray!50, fill=orange!10, rounded corners=3pt, minimum height=0.6cm, 
                          font=\ttfamily\footnotesize, text width=1.8cm, align=center},
        bucket_entry/.style={draw=gray!50, fill=orange!10, rounded corners=6pt, minimum height=0.6cm, 
                             font=\ttfamily\footnotesize, text width=2.5cm, align=center},  % Reduced width
        arrow/.style={->, >=stealth, thick, draw=gray!70},
        label/.style={font=\small\bfseries, align=center}
    ]
    
    % Define fixed positions for better spacing (same center as the first figure)
    \def\centerhorizontal{4.0}
    
    % Short Patterns Hash Map Title
    \node[label] at (\centerhorizontal, 3.5) {Short Patterns Hash Map};
    
    % Short Patterns Hash Map
    \node[map_key] (short1_key) at (\centerhorizontal-1.35, 2.8) {(\substring{hello}, 5)};
    \node[map_value] (short1_val) at (\centerhorizontal+1.5, 2.8) {723};
    \draw[arrow] (short1_key.east) -- (short1_val.west);
    
    \node[map_key] (short2_key) at (\centerhorizontal-1.35, 2.1) {(\substring{to}, 2)};
    \node[map_value] (short2_val) at (\centerhorizontal+1.5, 2.1) {312};
    \draw[arrow] (short2_key.east) -- (short2_val.west);
    
    \node[map_key] (short3_key) at (\centerhorizontal-1.35, 1.4) {(\substring{data}, 4)};
    \node[map_value] (short3_val) at (\centerhorizontal+1.5, 1.4) {459};
    \draw[arrow] (short3_key.east) -- (short3_val.west);
    
    % Long Patterns Bucket Title
    \node[label] at (\centerhorizontal, 0.3) {Long Patterns Buckets};
    
    % Centered Long Patterns Bucket Key (First Bucket)
    \node[map_key] (bucket1) at (\centerhorizontal-1.7, -0.4) {\substring{compress}};  % Moved closer to center
    
    % Connecting arrow from "compress" to the bucket
    \draw[arrow] (bucket1.south) -- +(0,-0.4);
    
    % Unified Bucket Vector (First Bucket)
    \node[bucket_entry] (long1) at (\centerhorizontal-1.7, -1.4) {(\substring{ion algo}, 8), 1499};
    \node[bucket_entry] (long2) at (\centerhorizontal-1.7, -2.1) {(\substring{ files}, 6), 944};
    \node[bucket_entry] (long3) at (\centerhorizontal-1.7, -2.8) {(\substring{or}, 2), 2245};
    
    % Centered Long Patterns Bucket Key (Second Bucket)
    \node[map_key] (bucket2) at (\centerhorizontal+1.7, -0.4) {\substring{database}};  % Moved closer to center
    
    % Connecting arrow from "encrypt" to the bucket
    \draw[arrow] (bucket2.south) -- +(0,-0.4);
    
    % Unified Bucket Vector (Second Bucket)
    \node[bucket_entry] (long4) at (\centerhorizontal+1.7, -1.4) {(\substring{ schema}, 7), 1782};
    \node[bucket_entry] (long5) at (\centerhorizontal+1.7, -2.1) {(\substring{ table}, 6), 2078};
    %\node[bucket_entry] (long6) at (\centerhorizontal+1.7, -2.8) {(\substring{ index}, 6), 1120};
    
    \end{tikzpicture}
    \caption{Visualization of the LPM data structures: Short patterns are stored in a hash map for direct lookups, while long patterns are grouped by an 8-byte prefix. The corresponding suffixes are arranged in descending length order. Each prefix maps to a dynamic bucket—a vector of suffixes—that supports insertions and length-based reordering.}
    \label{fig:lpm_structure}
\end{figure}

The LPM algorithm is implemented using a two-tier strategy to manage patterns of varying lengths:
\begin{itemize}
    \item \textit{Short patterns} ($\leq 8$ bytes) are stored in a hash map for direct lookup. The pattern data is packed into a 64-bit integer, allowing for efficient comparison and retrieval.
    \item \textit{Long patterns} ($>8$ bytes) are grouped by their 8-byte prefix. The suffixes of these patterns are stored contiguously in the same bucket, enabling efficient searching and matching.
\end{itemize}
As shown in \autoref{fig:lpm_structure}, the LPM algorithm utilizes two primary data structures to handle each category of patterns:
\begin{itemize}
    \item \textit{Hash Map for Short Patterns}: A hash map that stores short patterns. Each entry in the hash map is a key-value pair where the key is a tuple containing the 64-bit representation of the pattern and its length, and the value is the associated token ID.
    \item \textit{Buckets for Long Patterns}: A hash map that stores long patterns. Each entry in this hash map corresponds to an 8-byte prefix, and the value is a vector of tuples. Each tuple contains the suffix of the pattern, its length, and the associated token ID. The suffixes within each bucket are sorted in descending order so that the search can stop as soon as a match is found.
\end{itemize}

\subsubsection{Searching the Longest Match}
The process of identifying the longest possible dictionary match for a given position in the input data is provided in \autoref{alg:lpm_search}. The approach is divided into two main cases based on the remaining input length:
\begin{itemize}
    \item \textit{Long pattern matching} (lines 2-12): For input longer than 8 bytes, the algorithm:
    \begin{itemize}
        \item Uses the first 8 bytes as a key to locate a bucket of potential matches.
        \item Compares the subsequent bytes (suffix) with entries in the bucket.
        \item Returns immediately when it finds a match, since entries are sorted by descending length.
    \end{itemize}
    %\begin{enumerate}
%        \item \textit{Extract the 8-Byte Prefix}: The first 8 bytes of the input string are extracted and converted into a 64-bit integer. This prefix is used as a key to locate the corresponding bucket in the buckets hash map.
%        \item \textit{Extract the Suffix}: The remaining bytes of the input string (beyond the first 8 bytes) are also converted into a 64-bit integer, representing the suffix. The length of this suffix is calculated as the minimum of the remaining input length and 16 bytes (to handle cases where the input is longer than 16 bytes).
%        \item \textit{Search the Bucket}: Once the bucket corresponding to the 8-byte prefix is located, the algorithm iterates through the suffixes stored in the bucket. The suffixes are sorted in descending order of length, ensuring that the longest possible match is found first. Verifying whether a bucket entry matches the suffix takes constant time, using a single XOR operation.
%        \item \textit{Return the Match}: If a match is found, the algorithm returns the associated token ID and the total length of the match (8 bytes for the prefix plus the length of the matched suffix). If no match is found in the bucket, the algorithm proceeds to handle the input as a short pattern.
%\end{enumerate}
    \item \textit{Short pattern matching} (lines 13-19): For input of at most 8 bytes, the algorithm switches to a simpler strategy that directly searches the hash map for the longest possible match. The input bytes are packed into a 64-bit integer, and the algorithm iterates over possible prefix lengths, starting from the maximum possible length (8 bytes) and decreasing down to 1 byte. Since the dictionary is initialized with all the 1-byte values, it is guaranteed to find a match at this stage.
\end{itemize}

\begin{algorithm}
\begin{algorithmic}[1]
\Function{LPM\_Search}{\texttt{input}, \texttt{sp\_map}, \texttt{lp\_buckets}}
    \If{\texttt{len}(\texttt{input}) $>$ 8} 
        \State \texttt{prefix} $\gets$ \texttt{input}[:8]
        \If{\texttt{lp\_buckets}.\texttt{contains}(\texttt{prefix})}
            \State \texttt{bucket} $\gets$ \texttt{lp\_buckets}[\texttt{prefix}]
            \ForAll{(\texttt{suffix}, \texttt{token\_id}) \texttt{in} \texttt{bucket}}
                \If{\texttt{IsPrefix}(\texttt{input}[8:], \texttt{suffix})}
                    \State \Return (\texttt{token\_id}, 8 + \texttt{len}(\texttt{suffix}))
                \EndIf
            \EndFor
        \EndIf
    \EndIf
    \ForAll{\texttt{length} \texttt{in} \texttt{min}(\texttt{len}(\texttt{input}), 8) \texttt{to} 1}
        \State \texttt{prefix} $\gets$ \texttt{input}[:\texttt{length}]
        \If{\texttt{sp\_map}.\texttt{contains}(\texttt{prefix})}
            \State \texttt{token\_id} $\gets$ \texttt{sp\_map}[\texttt{prefix}]
            \State \Return (\texttt{token\_id}, \texttt{length})
        \EndIf
    \EndFor
\EndFunction
\end{algorithmic}
\caption{Longest Prefix Matching Search}
\label{alg:lpm_search}
\end{algorithm}

A key distinction between \compressor{\algorithmname{}} and \compressor{\algorithmname{16}} lies in how long pattern suffixes are handled, i.e., how we perform the \texttt{IsPrefix} check at line 7. In \compressor{\algorithmname{}}, suffixes of long patterns can exceed 8 bytes, which prevents packing them into a single machine word and requires more expensive memory accesses and comparisons during suffix matching. In contrast, \compressor{\algorithmname{16}} guarantees that all long pattern suffixes fit within 8 bytes. This enables each suffix to be packed into a single 64-bit integer, allowing highly efficient comparisons through bitwise operations such as XOR and trailing zero counting. This approach is shown in \autoref{alg:xor_prefix_check}, which computes the length of the shared prefix by counting the number of matching lowest-order bytes between two packed 64-bit values. The XOR operation identifies the differing bits (line 10), and then an efficient bitwise operation is used to efficiently determine how many full bytes match from the start (lines 11-12). Since 64-bit packed strings shorter than 8 bytes are padded with zeroes at the most significant end, the final comparison at line 6 ensures that only the actual prefix is considered, and not any artificial match caused by padding.

\begin{algorithm}
\begin{algorithmic}[1]
\Function{IsPrefix}{\texttt{input}, \texttt{prefix}}
    \If{\texttt{len}(\texttt{prefix}) > \texttt{len}(\texttt{input})}
        \State \Return \texttt{False}
    \EndIf
    \State \texttt{shared} $\gets$ \Call{SharedPrefixSize}{\texttt{input}, \texttt{prefix}}
    \State \Return \texttt{shared} $\geq$ \texttt{len}(\texttt{prefix})
\EndFunction
\\
\Function{SharedPrefixSize}{\texttt{s1}, \texttt{s2}}
    \State \texttt{diff} $\gets$ \texttt{s1} $\oplus$ \texttt{s2}
    \State \texttt{zeros} $\gets$ \texttt{count\_trailing\_zeros}(\texttt{diff})
    \State \Return \texttt{zeros} / 8
\EndFunction
\end{algorithmic}
\caption{XOR-based prefix comparison for 64-bit packed strings}
\label{alg:xor_prefix_check}
\end{algorithm}

\subsubsection{Static Longest Prefix Matching} \label{sec:static_lpm} 
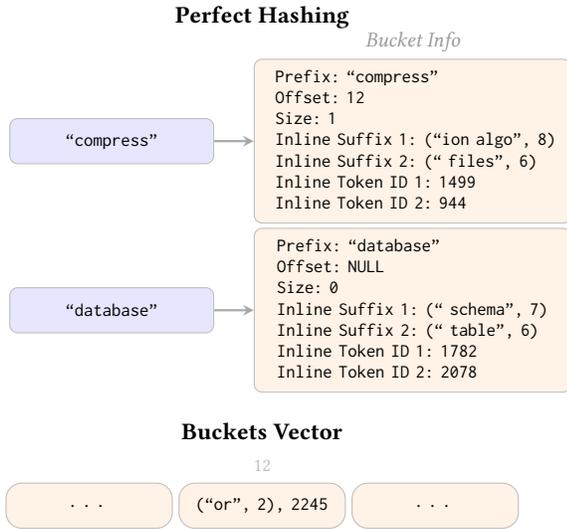
\begin{figure}
    \centering
    \begin{tikzpicture}[
        map_key/.style={draw=gray!50, fill=blue!10, rounded corners=3pt, minimum height=0.6cm, 
                        font=\ttfamily\footnotesize, text width=2.5cm, align=center},
        map_value/.style={draw=gray!50, fill=orange!10, rounded corners=3pt, minimum height=0.6cm, 
                          font=\ttfamily\footnotesize, text width=4cm, align=center},  % Increased width
        bucket_entry/.style={draw=gray!50, fill=orange!10, rounded corners=6pt, minimum height=0.6cm, 
                             font=\ttfamily\footnotesize, text width=2.0cm, align=center},
        arrow/.style={->, >=stealth, thick, draw=gray!70},
        label/.style={font=\normalsize\bfseries, align=center},
        annotation/.style={font=\small\itshape, text=gray!80},
        index/.style={font=\footnotesize, text=gray!50, above, yshift=1pt}
    ]

    % Define fixed positions for better spacing
    \def\centerhorizontal{4.0}
    
    % Perfect Hashing Title (moved higher)
    \node[label] at (\centerhorizontal, 6.0) {Perfect Hashing};
    
    % First Perfect Hashing Example: "compress"
    \node[map_key] (prefix1) at (\centerhorizontal-2.0, 4.35) {\substring{compress}};  % Moved left
    \node[map_value] (bucket_info1) at (\centerhorizontal+2.0, 4.35) {  % Moved right
        \begin{tabular}{l}
            Prefix: \substring{compress}\\
            Offset: 12 \\
            Size: 1 \\
            Inline Suffix 1: (\substring{ion algo}, 8) \\
            Inline Suffix 2: (\substring{ files}, 6) \\
            Inline Token ID 1: 1499\\
            Inline Token ID 2: 944\\
        \end{tabular}
    };
    \draw[arrow] (prefix1.east) -- (bucket_info1.west);  % Arrow between boxes
    
    % Label for Bucket Info Structure (non-bold)
    \node[annotation] at (\centerhorizontal+2.0, 5.67) {Bucket Info};
    
    % Second Perfect Hashing Example: "database"
    \node[map_key] (prefix2) at (\centerhorizontal-2.0, 2.1) {\substring{database}};  % Moved left
    \node[map_value] (bucket_info2) at (\centerhorizontal+2.0, 2.1) {  % Moved right
        \begin{tabular}{l}
            Prefix: \substring{database}\\
            Offset: NULL \\
            Size: 0 \\
            Inline Suffix 1: (\substring{ schema}, 7) \\
            Inline Suffix 2: (\substring{ table}, 6) \\
            Inline Token ID 1: 1782\\
            Inline Token ID 2: 2078\\
        \end{tabular}
    };
    \draw[arrow] (prefix2.east) -- (bucket_info2.west);  % Arrow between boxes

    % External Buckets Title
    \node[label] at (\centerhorizontal, 0.5) {Buckets Vector};
    
    % External Buckets Example
    \node[bucket_entry] (external2) at (\centerhorizontal-2.3, -0.5) {\dots};
    
    \node[bucket_entry] (external2) at (\centerhorizontal, -0.5) {(\substring{or}, 2), 2245};
    \node[index] at (external2.north) {12};  

    \node[bucket_entry] (external3) at (\centerhorizontal+2.3, -0.5) {\dots};
    
    \end{tikzpicture}
    \caption{Visualization of \compressor{\algorithmname{16}}'s Static LPM data structures for long patterns. A perfect hash function maps prefixes directly to the bucket info structure, which fits in a single cache line. This structure contains the prefix, offset, size, and inline suffixes. Additional suffixes are stored in an external vector, accessed via the offset and size stored in the bucket info.}
    \label{fig:static_lpm_structure}
\end{figure}

At the end of the training phase, the LPM data structures can be processed to exploit the static nature of the parsing stage. As shown in \autoref{fig:static_lpm_structure}, the static version of the LPM algorithm leverages perfect hashing to achieve efficient lookups for long patterns. This optimization specifically applies to \compressor{\algorithmname{16}}, as the upper bound on the maximum token length allows the first few suffixes to be inlined directly into the lookup structure.

The transition from the dynamic to the static version involves a finalization step, where the data structures are reorganized to optimize for read-only access. The finalization process is as follows:
\begin{itemize}
    \item \textit{Perfect Hashing for Long Patterns}: The prefixes of long patterns are used as keys in a perfect hash function. This ensures that each prefix maps to a unique index in a precomputed table, eliminating collisions and reducing lookup time to constant complexity. The perfect hash function returns an index into a vector of bucket information, where each entry contains metadata about the corresponding bucket. This bucket information structure is designed to fit within a single cache line (64 bytes), ensuring efficient memory access.
    \item \textit{Inline Suffixes}: To further optimize memory access, the first few suffixes are stored inline directly in the bucket information structure. This avoids the need for separate memory accesses for small buckets, which are common in practice (see \autoref{fig:bucket_len_distribution}). When a lookup occurs, the algorithm retrieves the bucket index using the perfect hash function and checks the inline suffixes first. If no match is found, a fallback mechanism handles the cases where more suffixes exist.
    \item \textit{Buckets Organization}: For buckets that contain more entries than the inline storage capacity, the additional suffixes are stored in a contiguous vector. The bucket information structure includes an offset into this vector, together with the bucket size. As before, suffixes within each bucket are sorted in descending order of length, ensuring that the longest possible match is found first.
\end{itemize}

\subsubsection{Bounding Bucket Size}
\begin{figure}
  \centering
  \includegraphics[width=\linewidth]{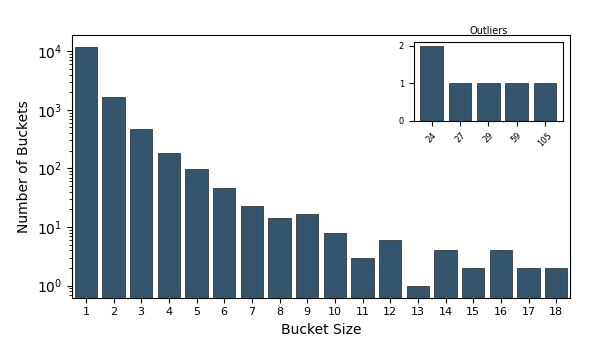}
  \caption{Distribution of \compressor{\algorithmname{16}}'s bucket sizes for the \dataset{Book Titles} dataset. The main plot shows buckets containing 1-18 suffixes, while the inset displays rare outliers with larger bucket sizes. Most buckets contain only a few suffixes: over 98\% of buckets contain 4 or fewer suffixes, with the majority being singleton buckets.}
  \label{fig:bucket_len_distribution}
\end{figure}

During longest prefix matching, long patterns (i.e., entries exceeding 8 bytes) are organized into buckets keyed by their 8-byte prefix. Each bucket contains the suffix of all patterns sharing the same prefix, sorted by descending length. Real-world datasets exhibit a heavy skew in bucket sizes: \autoref{fig:bucket_len_distribution} shows that for the \dataset{Book Titles} dataset using \compressor{\algorithmname{16}}, over 98\% of buckets contain at most 4 suffixes. This distribution holds across all tested datasets, confirming that most buckets require minimal search depth.

However, edge cases emerge when strings share long common prefixes, potentially resulting in buckets containing thousands of suffixes. While these large buckets are rare, they disproportionately impact performance: longer suffixes in these buckets typically represent minor variations of high-frequency prefixes, such as \texttt{customer\_id\_001} and \texttt{customer\_id\_002}. These niche variants contribute marginally to overall compression but incur linear search costs.

To bound worst‑case lookup costs, \compressor{\algorithmname{16}} limits each bucket to at most 128 suffixes, striking a practical balance between compression effectiveness and query efficiency. The unbounded variant \compressor{\algorithmname{}} is retained to illustrate the upper bound on compression achievable by our approach.

\subsection{Decompression}
The decompression process reconstructs the original data by expanding each 2-byte token ID using the dictionary built during the training phase. Each string is compressed independently, meaning that decompression can be performed on a per-string basis without scanning or decoding surrounding data. This allows fine-grained random access, as the decoder can decompress any individual string in isolation.

\begin{figure}
\centering
\begin{tikzpicture}[
    dictbox/.style={
        draw=gray!50, 
        minimum height=0.6cm,
        minimum width=0.6cm,
        anchor=south west, 
        fill=white, 
        line width=0.5pt,
        font=\ttfamily\footnotesize
    },
    offsetbox/.style={
        draw=gray!50, 
        minimum height=0.7cm,
        minimum width=0.7cm,
        font=\ttfamily\footnotesize, 
        align=center,
        fill=orange!10,
        rounded corners=2pt
    },
    arrow/.style={
        ->, 
        >=stealth, 
        thick, 
        draw=gray!70,
    },
    label/.style={
        font=\small\bfseries,
        anchor=east,
        inner xsep=0pt  % Reduce padding around label text
    },
    index/.style={
        font=\footnotesize,
        text=gray!50,
        above,
        yshift=1pt
    }
]

% Dictionary Array - moved label closer and aligned better
\node[label] at (-0.8, 0.3) {Data:};  % Moved from -1.2 to -0.8
\node[dictbox] (prev) at (0,0) {\tiny$\cdots$};
\foreach \x/\c [count=\i from 0] in {0/a,1/b,2/r,3/a,4/c,5/a,6/d} {
    \node[dictbox] (a\i) at (0.6+\x*0.6,0) {\c};
}
\node[dictbox] (next) at (4.8,0) {y};
\node[dictbox] (dots) at (5.4,0) {\tiny$\cdots$};

% Position indexes
\node[index] at (a0.north) {3769};
\node[index] at (next.north) {3776};

% Offsets Array - adjusted vertical spacing
\node[label] at (-0.8, -1.0) {Offsets:};  % Moved from -1.2 to -0.8
\node[offsetbox] (o0) at (0,-1.0) {\tiny$\cdots$};
\node[offsetbox] (o1) at (0.7,-1.0) {3769};
\node[offsetbox] (o2) at (1.4,-1.0) {3776};
\node[offsetbox] (dots) at (2.1,-1.0) {\tiny$\cdots$};

% Offset indexes
\node[index] at (o1.north) {793};
\node[index] at (o2.north) {794};

% Arrows
\draw[arrow] (o1.north) -- (a0.south);
\draw[arrow] (o2.north) -- (next.south);
\end{tikzpicture}
\caption{Memory layout of the dictionary data structures. The token with ID 793, storing the string \substring{abracad}, starts at position 3769, while the difference between consecutive offsets (793 and 794) implicitly encodes its length.}
\label{fig:dictionary_layout}
\end{figure}
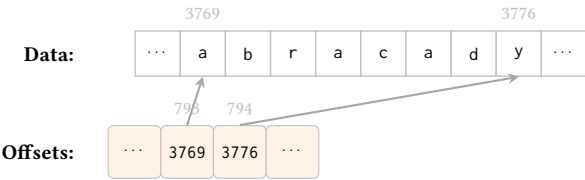

Since each token is represented as a fixed-size 2-byte value, the decompression process consists of simple and predictable lookup operations. To ensure efficient access, the dictionary is stored in a contiguous memory block, and each entry is indexed via an offset table (see \autoref{fig:dictionary_layout}). Given a token ID $i$, the corresponding entry can be retrieved from the dictionary by reading the bytes between $\texttt{offsets}[i]$ and $\texttt{offsets}[i+1]$. 

In the case of \compressor{\algorithmname{16}}, where all dictionary entries are guaranteed to be at most 16 bytes long, the decoder performs a fixed-size copy of 16 bytes for every token. This predictable memory access pattern enables efficient vectorized decoding, as modern CPUs can leverage SIMD instructions to copy 16 bytes in a single instruction. After the copy, the decoder simply advances the output buffer by the actual length of the token.

In the case of \compressor{\algorithmname{}} (see \autoref{alg:token_decompression}), where token lengths are unbounded, we use a two-stage decoding strategy optimized for the common case. The decoder copies the first 16 bytes unconditionally and performs an additional copy only if the token exceeds this length. This approach exploits the skewed token length distribution observed in most datasets (see \autoref{fig:token_length_distribution_book_titles_onpair}), where most patterns fit within 16 bytes.

\begin{algorithm}
\begin{algorithmic}[1]
\Function{decompress\_token}{\texttt{token\_id}, \texttt{data}, \texttt{offsets}, \texttt{buff}}
    \State \texttt{start} $\gets$ \texttt{offsets}[\texttt{token\_id}]
    \State \texttt{end} $\gets$ \texttt{offsets}[\texttt{token\_id} + 1]
    \State \texttt{length} $\gets$ \texttt{end} - \texttt{start}
    
    \State \texttt{copy}(\texttt{buff}, \texttt{data} + \texttt{start}, 16) \Comment{Exploit SIMD}
    
    \If{\texttt{length} > 16}
        \State \texttt{copy}(\texttt{buff} + 16, \texttt{data} + \texttt{start} + 16, \texttt{length} - 16) 
    \EndIf
    
    \State \Return \texttt{length}
\EndFunction
\end{algorithmic}
\caption{Fast unbounded token decompression}
\label{alg:token_decompression}
\end{algorithm}

\subsection{Dictionary Size Trade-Offs}
\begin{figure}
  \centering
  \includegraphics[width=\linewidth]{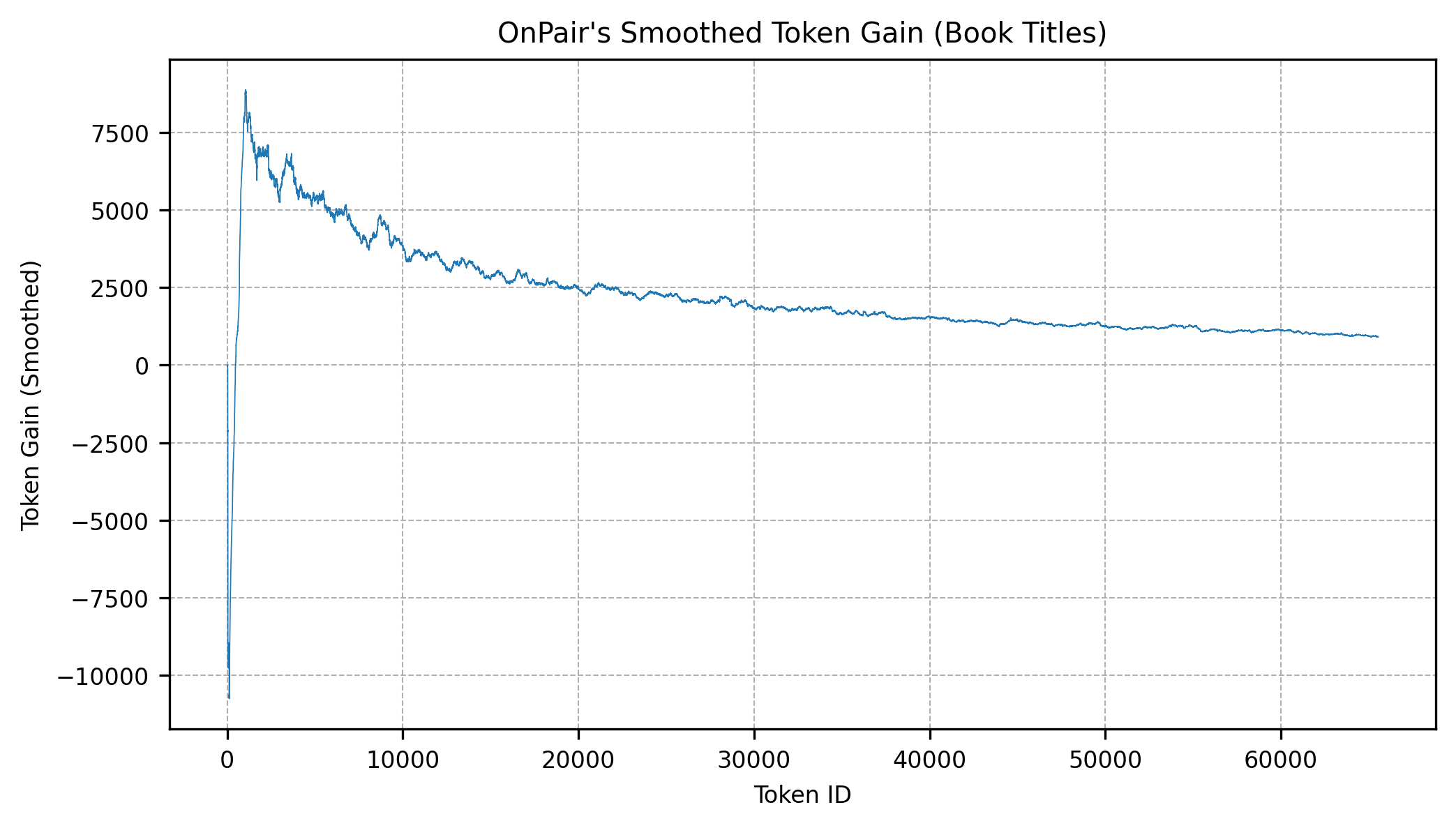}
  \caption{Smoothed token gain for \compressor{\algorithmname{}} on the \dataset{Book Titles} dataset. A moving average is applied to reveal overall trends using a window size of 655, which is approximately 1\% of the dictionary size. Initial tokens (IDs 0–255) represent single bytes and exhibit negative gain due to the 2-byte token ID overhead. Later tokens demonstrate diminishing returns, with smoothed gain decreasing as token ID increases, confirming that early dictionary entries capture the most valuable compression patterns.}
  \label{fig:smoothed_token_gain_book_titles_onpair}
\end{figure}

The dictionary size—determined by the number of bits per token—represents a fundamental compression parameter that requires careful balancing of competing objectives. While larger dictionaries can improve compression by capturing longer patterns, this benefit follows diminishing returns. As shown in \autoref{fig:smoothed_token_gain_book_titles_onpair}, early token additions contribute disproportionately to compression gain by encoding high-frequency substrings, while later additions target increasingly rare patterns with smaller space savings. Beyond the optimal size, the fixed per-token storage overhead outweighs the marginal compression benefit of new entries, ultimately reducing overall compression effectiveness.

The impact of dictionary size extends beyond compression ratios. Larger dictionaries yield longer average token lengths, which reduces the number of lookups required during decompression. However, this advantage is counterbalanced by more complex longest prefix matching and by memory hierarchy effects as bigger dictionaries spill into slower cache levels. Conversely, smaller dictionaries are faster to build and parse fewer pairs, but at the cost of shorter tokens and lower compression ratios.

\begin{table*}
  \caption{Compression performance of \compressor{\algorithmname{}} on the \dataset{Book Titles} dataset with varying dictionary sizes. The ``Bits'' column indicates the number of bits per token. Decompression speed measures throughput when decoding the full dataset from start to finish. Access reports the average time to retrieve a string from 1 million random point queries. Average token length refers to the compressed output. \compressor{\algorithmname{}}'s threshold is set to its minimum value to ensure that large dictionaries are fully populated.}
  \label{tab:bits_per_token}
  \begin{tabular}{rrrrrrr}
    \toprule
    Bits & Comp. Ratio & Comp. (MiB/s) & Decomp. (MiB/s) & Access (ns) & Dictionary (MiB) & Token Len. (B) \\
    \midrule
    \rowcolor{gray!0}  9 & 1.522 &  85 & 2\,819 & 277 &  0.003 &  1.71 \\
    \rowcolor{gray!5} 10 & 1.693 &  97 & 3\,442 & 270 &  0.007 &  2.12 \\
    \rowcolor{gray!0} 11 & 1.827 & 109 & 3\,699 & 267 &  0.015 &  2.51\\
    \rowcolor{gray!5} 12 & 1.983 & 127 & 4\,607 & 265 &  0.032 &  2.98 \\
    \rowcolor{gray!0} 13 & 2.153 & 139 & 4\,491 & 263 &  0.070 &  3.50 \\
    \rowcolor{gray!5} 14 & 2.330 &  75 & 4\,653 & 262 &  0.152 &  4.09 \\
    \rowcolor{gray!0} 15 & 2.512 & 112 & 4\,546 & 263 &  0.333 &  4.73 \\
    \rowcolor{gray!5} 16 & 2.682 & 135 & 5\,509 & 263 &  0.727 &  5.41 \\
    \rowcolor{gray!0} 17 & 2.839 &  99 & 4\,000 & 275 &  1.593 &  6.16 \\
    \rowcolor{gray!5} 18 & 2.952 &  62 & 3\,105 & 287 &  3.504 &  6.97 \\
    \rowcolor{gray!0} 19 & 2.971 &  31 & 2\,485 & 316 &  7.729 &  7.88 \\
    \rowcolor{gray!5} 20 & 2.809 &  13 & 1\,988 & 373 & 17.130 &  8.98 \\
    \rowcolor{gray!0} 21 & 2.370 &   5 & 1\,046 & 438 & 38.050 & 10.52 \\
  \bottomrule
\end{tabular}
\end{table*}

Ultimately, the best dictionary size depends on the dataset and the specific performance goals. However, across our experiments, we found that allocating 16 bits per token strikes a practical and robust compromise between these factors. This choice avoids the overhead of bit packing and misaligned reads while offering a strong balance of compression ratio and speed. \autoref{tab:bits_per_token} reports performance metrics of \compressor{\algorithmname{}} on the \dataset{Book Titles} dataset for different dictionary sizes, as determined by the number of bits per token. The compression ratio increases steadily as the number of available tokens grows, peaking around 19 bits. However, this improvement comes at a steep cost: both compression and decompression speeds degrade significantly beyond 16 bits.

\section{Experiments}
\label{sec:experiments}
\subsection{Code Availability}
We provide custom benchmarking code tailored to this compression task, implemented in both C++ and Rust. The C++ benchmark is used to evaluate native compression algorithms (\compressor{FSST}, \compressor{LZ4}, and \compressor{Zstandard}), while the Rust benchmark is used for the remaining methods. The static longest prefix matching algorithm in \compressor{\algorithmname{16}} relies on \textit{PtrHash} \cite{koerkamp2025ptrhash}, a (minimal) perfect hash function not readily available in C++. The source code is publicly available at:
\begin{center}
    \href{https://github.com/gargiulofrancesco/compression_benchmark_cpp}{github.com/gargiulofrancesco/compression\_benchmark\_cpp}\\
    \href{https://github.com/gargiulofrancesco/compression_benchmark_rs}{github.com/gargiulofrancesco/compression\_benchmark\_rs}
\end{center}

\subsection{Hardware and Software Environment}
Experiments were performed on a desktop system equipped with an Intel Core Ultra 7 265K processor and 128~GB of RAM. The CPU employs a hybrid architecture comprising 20 cores (8 performance cores and 12 efficiency cores) and 20 threads. To ensure consistent and comparable single-thread performance, each test was executed on a single performance core, explicitly pinned to that core.

The performance cores support turbo frequencies up to 5.5 GHz and are equipped with 64 KiB of L1 instruction cache, 48 KiB of L1 data cache, and 3 MiB of private L2 cache. Additionally, the processor contains 30 MiB of shared L3 cache. The processor supports SIMD instruction sets up to and including AVX2.

The software environment consisted of Ubuntu 25.04 (64-bit) running the Linux kernel version 6.14.0-15-generic. All code was compiled using the following toolchains and options:
\begin{itemize}
    \item Rust: rustc 1.87.0 and Cargo 1.87.0, built with the ``\textnormal{-}\textnormal{-}\texttt{release}'' flag and ``\texttt{-C target-cpu=native}'' for architecture-specific optimizations.
    \item C++: g++ 14.2.0, using ``\texttt{-O3} \texttt{-march=native}'' to enable architecture-specific optimizations.
\end{itemize}
Note that the evaluation machine does not support AVX‑512 instructions; \compressor{FSST} results are therefore reported without its AVX‑512 optimizations, which would otherwise further accelerate compression speed.

\subsection{Datasets}
The experiments were conducted on a collection of real-world textual datasets, selected to encompass a range of characteristics, including varying string lengths, total sizes, and content types. In the following, we provide a brief overview of each dataset.
\begin{itemize}
    \item \dataset{Book Reviews} and \dataset{Book Titles} \cite{hou2024retrieval}: Product titles and reviews sourced from the Amazon Books section, with the first 500~MiB of reviews selected.
    \item \dataset{News Headlines} \cite{kulkarni2018headlines}: ABC News headlines, collected over 19 years.
    \item \dataset{Tweets} \cite{sahni2017twitter}: A collection of tweets annotated for sentiment analysis.
    \item \dataset{URLs} \cite{gokaslan2019openweb}: A large corpus of URLs, curated to replicate the quality of web content found in OpenAI's WebText.
\end{itemize}
\autoref{tab:dataset_overview} provides a summary of the datasets used in our experiments, including their size, number of rows, and average entry length.
\begin{table}
  \caption{General information about datasets used in experiments.}
  \label{tab:dataset_overview}
  \begin{tabular}{lrrr}
    \toprule
    Dataset&Size (MiB)&Rows&Avg. Length (B)\\
    \midrule
    \rowcolor{gray!0} \dataset{Book Reviews} & 498.10 & 1\,243\,977 & 420\\
    \rowcolor{gray!5} \dataset{Book Titles} & 220.59 & 4\,448\,181 & 52\\
    \rowcolor{gray!0} \dataset{News Headlines} & 48.99 & 1\,244\,184 & 41\\
    \rowcolor{gray!5} \dataset{Tweets} & 113.06 & 1\,600\,000 & 74\\
    \rowcolor{gray!0} \dataset{URLs} & 1\,846.17 & 23\,031\,394 & 84\\
  \bottomrule
\end{tabular}
\end{table}

\subsection{Benchmark Results}
We evaluate the performance of \compressor{\algorithmname{}} and \compressor{\algorithmname{16}} against several state-of-the-art compression algorithms across four key metrics: \textit{compression ratio}, \textit{compression speed}, \textit{decompression speed}, and \textit{average random access time}. Our experimental setup distinguishes between two categories of compressors based on their access granularity:
\begin{itemize}
    \item \textit{Block-Based Compressors} (\compressor{LZ4}, \compressor{Zstandard}) group strings into fixed-size blocks of 64~KiB before compression, allowing them to exploit redundancy across multiple strings within each block. To support random access, we implement a simple block-level cache: when a string is requested, the entire 64~KiB block containing it is decompressed and stored in memory. If subsequent queries access strings from the same block, decompression is skipped and results are served directly from the cache.
    \item \textit{Field-Level Compressors} (\compressor{BPE}, \compressor{FSST}, \compressor{\algorithmname{}}, \compressor{\algorithmname{16}}) compress each string independently using an external dictionary, enabling direct random access without block-level overhead. These methods trade some compression efficiency for faster fine-grained access patterns. % \compressor{BPE} implements the same token decoding routine as \compressor{\algorithmname{}}, as detailed in \autoref{alg:token_decompression}, and thus their decoding performance are expected to be very similar.
\end{itemize}
In addition, we include \compressor{Raw}, an uncompressed baseline that stores strings in their original form. Since no compression or decoding is performed, \compressor{Raw} has no overhead and serves as a neutral reference for evaluating the trade‑offs introduced by compression.

\autoref{tab:final_benchmark_results} summarizes performance across all datasets. Random access latency is measured as the average time to retrieve and decompress individual strings over 1 million uniformly distributed queries, while decompression throughput reflects decoding the entire dataset sequentially. Several observations arise from these results:

\begin{itemize}
    \item \textit{Compression Ratio}: \compressor{\algorithmname{}} and \compressor{\algorithmname{16}} achieve compression levels close to \compressor{BPE}, while consistently outperforming \compressor{FSST}. Block‑based compressors yield noticeably lower ratios, especially \compressor{Zstandard}, which typically excels on large contiguous data. This gap indicates that our datasets exhibit limited cross‑string redundancy, making field‑level approaches with external dictionaries more effective.
    \item \textit{Compression Speed}: Among field‑level compressors, \compressor{FSST} achieves the highest compression throughput, even without its AVX‑512‑accelerated implementation (unavailable on our evaluation hardware). \compressor{\algorithmname{16}} narrows the gap, providing a substantially better compression ratio while maintaining competitive throughput. \compressor{\algorithmname{}} sacrifices some speed for slightly higher compression ratios due to its more complex longest prefix matching.
    \item \textit{Decompression Speed}: \compressor{\algorithmname{16}} achieves the highest decompression speed across all datasets, thanks to bounded token length, longer average tokens, and SIMD‑friendly decoding. \compressor{FSST}, \compressor{LZ4} and \compressor{\algorithmname{}} also offer strong performance, while \compressor{Zstandard} lags behind due to heavier entropy coding. 
    \item \textit{Average Random Access Time}: Block-based methods suffer under uniformly distributed queries, as each access triggers full block decompression with little opportunity for reuse. In contrast, field-level compressors enable fast point-wise queries by compressing strings independently, and can even outperform \compressor{Raw} due to better cache efficiency. Their smaller memory footprints fit more easily within CPU cache hierarchies, reducing cache misses and memory bandwidth pressure. This advantage grows under high query loads, where compressed data exhibits better spatial locality. It is particularly evident on the smallest dataset, \dataset{News Headlines}, whose uncompressed size exceeds the machine's L3 cache, while the compressed version fits entirely within it—substantially reducing cache misses.
\end{itemize}

\begin{table*}
  \caption{Compression ratio, (de)compression speed, and average random access latency across all datasets. Decompression speed is measured over the full dataset from start to finish. \compressor{Raw} serves as an uncompressed baseline with no decompression overhead. Block-based compressors (\compressor{LZ4} and \compressor{Zstandard}) group strings together in blocks of 64~KiB to achieve higher compression ratios but suffer from slower random access due to full-block decompression. In contrast, \compressor{FSST}, \compressor{\algorithmname{}}, and \compressor{\algorithmname{16}} operate on individual strings, enabling much faster point-wise queries. \compressor{FSST} compression speed results are reported without AVX‑512 optimizations due to hardware limitations.}
  \label{tab:final_benchmark_results}
  \begin{tabular}{llrrrr}
    \toprule
    Dataset&Compressor&C. Ratio&Comp. (MiB/s)&Decomp. (MiB/s)&Access (ns)\\
    \midrule            
    \rowcolor{gray!25} \dataset{Book Reviews} & & & & & \\
        \rowcolor{gray!0} & \compressor{Raw}                & 1.00 & 22\,550 & 22\,190 & 294 \\
        \rowcolor{gray!5} & \compressor{LZ4}                & 1.64 & 498 & 3\,754 & 15\,712 \\
        \rowcolor{gray!0} & \compressor{Zstandard}          & 2.53 & 314 & 1\,663 & 36\,917 \\
        \rowcolor{gray!5} & \compressor{BPE}                & 3.47 & 2 & 6\,835 & 313  \\
        \rowcolor{gray!0} & \compressor{FSST}               & 1.89 & 479 & 5\,657 & 333 \\
        \rowcolor{gray!5} & \compressor{\algorithmname{}}   & 3.30 & 131 & 6\,658 & 315 \\
        \rowcolor{gray!0} & \compressor{\algorithmname{16}} & 3.28 & 229 & 7\,776 & 308 \\
        % \rowcolor{gray!0} & \compressor{Sampled BPE}        & 3.44 & 45 & 6\,679 & 316 \\
        % \rowcolor{gray!5} & \compressor{Sampled BPE16}      & 3.42 &  52 & 8\,126 & 303 \\
                             
    \rowcolor{gray!25} \dataset{Book Titles} & & & & & \\    
        \rowcolor{gray!0} & \compressor{Raw}                & 1.00 & 19\,486 & 22\,664 & 268 \\
        \rowcolor{gray!5} & \compressor{LZ4}                & 1.42 & 425 & 4\,067 & 16\,158 \\
        \rowcolor{gray!0} & \compressor{Zstandard}          & 2.04 & 283 & 1\,543 & 39\,848 \\
        \rowcolor{gray!5} & \compressor{BPE}                & 2.84 & 3 & 5\,790 & 259 \\
        \rowcolor{gray!0} & \compressor{FSST}               & 1.68 & 325 & 4\,964 & 269 \\
        \rowcolor{gray!5} & \compressor{\algorithmname{}}   & 2.83 & 114 & 5\,611 & 260 \\
        \rowcolor{gray!0} & \compressor{\algorithmname{16}} & 2.81 & 177 & 6\,732 & 261 \\
        % \rowcolor{gray!0} & \compressor{Sampled BPE}        & 2.81 & 36 & 5\,615 & 262 \\
        % \rowcolor{gray!5} & \compressor{Sampled BPE16}      & 2.80 &  41 & 6\,725 & 262 \\
                             
    \rowcolor{gray!25} \dataset{News Headlines}  & & & & & \\   
        \rowcolor{gray!0} & \compressor{Raw}                & 1.00 & 16\,737 & 22\,335 & 195 \\
        \rowcolor{gray!5} & \compressor{LZ4}                & 1.53 & 452 & 3\,780 & 15\,010 \\
        \rowcolor{gray!0} & \compressor{Zstandard}          & 2.41 & 302 & 1\,698 & 36\,120 \\
        \rowcolor{gray!5} & \compressor{BPE}                & 3.37 & 3 & 6\,352 & 152 \\
        \rowcolor{gray!0} & \compressor{FSST}               & 1.89 & 403 & 5\,683 & 175 \\
        \rowcolor{gray!5} & \compressor{\algorithmname{}}   & 3.25 & 100 & 6\,220 & 154 \\
        \rowcolor{gray!0} & \compressor{\algorithmname{16}} & 3.23 & 137 & 7\,460 & 156 \\
        % \rowcolor{gray!0} & \compressor{Sampled BPE}        & 3.38 & 12 & 6\,279 & 156 \\
        % \rowcolor{gray!5} & \compressor{Sampled BPE16}      & 3.33 &  13 & 7\,878 & 157 \\
                             
    \rowcolor{gray!25} \dataset{Tweets}  & & & & & \\   
        \rowcolor{gray!0} & \compressor{Raw}                & 1.00 & 19\,017 & 22\,517 & 239 \\
        \rowcolor{gray!5} & \compressor{LZ4}                & 1.41 & 417 & 3\,752 & 16\,664 \\
        \rowcolor{gray!0} & \compressor{Zstandard}          & 2.07 & 282 & 1\,544 & 39\,984 \\ 
        \rowcolor{gray!5} & \compressor{BPE}                & 2.65 & 3 & 6\,119 & 225 \\
        \rowcolor{gray!0} & \compressor{FSST}               & 1.74 & 442 & 5\,207 & 240 \\
        \rowcolor{gray!5} & \compressor{\algorithmname{}}   & 2.65 & 99 & 5\,791 & 225 \\
        \rowcolor{gray!0} & \compressor{\algorithmname{16}} & 2.64 & 155 & 6\,502 & 226 \\
        % \rowcolor{gray!0} & \compressor{Sampled BPE}        & 2.69 & 22 & 5\,959 & 228 \\
        % \rowcolor{gray!5} & \compressor{Sampled BPE16}      & 2.68 &  23 & 6\,715 & 226\\
                                
    \rowcolor{gray!25} \dataset{URLs}  & & & & & \\     
        \rowcolor{gray!0} & \compressor{Raw}                & 1.00 & 21\,142 & 22\,538 & 355 \\
        \rowcolor{gray!5} & \compressor{LZ4}                & 1.41 & 474 & 4\,392 & 15\,632 \\
        \rowcolor{gray!0} & \compressor{Zstandard}      & 2.05 & 306 & 1\,598 & 38\,901 \\ 
        \rowcolor{gray!5} & \compressor{BPE}                & 2.93 & 2 & 4\,575 & 340 \\
        \rowcolor{gray!0} & \compressor{FSST}               & 1.68 & 504 & 5\,175 & 359 \\
        \rowcolor{gray!5} & \compressor{\algorithmname{}}   & 2.80 & 14 & 4\,353 & 341 \\
        \rowcolor{gray!0} & \compressor{\algorithmname{16}} & 2.68 & 186 & 6\,623 & 335 \\
        % \rowcolor{gray!0} & \compressor{Sampled BPE}        & 2.93 & 11 & 4\,355 & 339 \\
        % \rowcolor{gray!5} & \compressor{Sampled BPE16}      & 2.73 & 113 & 6\,777 & 334 \\
                              
  \bottomrule
\end{tabular}
\end{table*}

\subsection{Dictionary Memory Footprint}
\label{sec:dictionary_memory_footprint}
We examine the memory footprint of the dictionaries produced by \compressor{\algorithmname{}} and \compressor{\algorithmname{16}} across datasets. Each dictionary consists of two components: a contiguous data region storing the raw bytes of each dictionary entry, and an offset array used to locate token boundaries during decoding. Offsets are stored as 4-byte integers and occupy up to 0.25~MiB (in case the dictionary is full). The remaining portion of the dictionary memory is used for the actual token data.

\autoref{tab:dictionary_size} reports the total size of the dictionary as well as the portion used for the token data alone. As expected, \compressor{\algorithmname{16}} produces slightly more compact dictionaries compared to \compressor{\algorithmname{}}, due to the maximum entry length constraint. However, the difference is modest and, in all cases, the total memory usage remains well under 1~MiB.

\begin{table}
  \caption{Dictionary memory footprint in MiB for \compressor{\algorithmname{}} and \compressor{\algorithmname{16}} across datasets. Each configuration allocates 0.25~MiB for the offset array, while the remaining space is used to store the dictionary data.}
  
  \label{tab:dictionary_size}
  \centering
  \begin{tabular}{llrr}
    \toprule
    Compressor & Dataset & Total (MiB) & Data (MiB) \\
    \midrule
    \rowcolor{gray!25} \compressor{\algorithmname{}} & & & \\
                           \rowcolor{gray!0} & \dataset{Book Reviews}    & 0.762 & 0.512\\
                           \rowcolor{gray!5} & \dataset{Book Titles}     & 0.708 & 0.458\\
                           \rowcolor{gray!0} & \dataset{News Headlines}  & 0.850 & 0.600\\
                           \rowcolor{gray!5} & \dataset{Tweets}          & 0.690 & 0.440\\
                           \rowcolor{gray!0} & \dataset{URLs}            & 0.746 & 0.496\\
    \rowcolor{gray!25} \compressor{\algorithmname{16}} & & & \\
                           \rowcolor{gray!0} & \dataset{Book Reviews}    & 0.744 & 0.494\\
                           \rowcolor{gray!5} & \dataset{Book Titles}     & 0.688 & 0.438\\
                           \rowcolor{gray!0} & \dataset{News Headlines}  & 0.776 & 0.526\\
                           \rowcolor{gray!5} & \dataset{Tweets}          & 0.680 & 0.430\\
                           \rowcolor{gray!0} & \dataset{URLs}            & 0.658 & 0.408\\
    \bottomrule
  \end{tabular}
\end{table}

\subsection{Decompression Speed}
We highlight two key observations that help explain \compressor{\algorithmname{16}}'s decompression performance: (1) the average token length, which influences the number of required dictionary lookups, and (2) the skewness of token frequencies, which affects cache efficiency during these lookups. These insights shed light on how \compressor{\algorithmname{16}} maintains competitive decompression speed despite its larger dictionary size compared to \compressor{FSST}, as shown in \autoref{tab:final_benchmark_results}.

\subsubsection{Average Token Length}
\begin{figure}
  \centering
  \includegraphics[width=\linewidth]{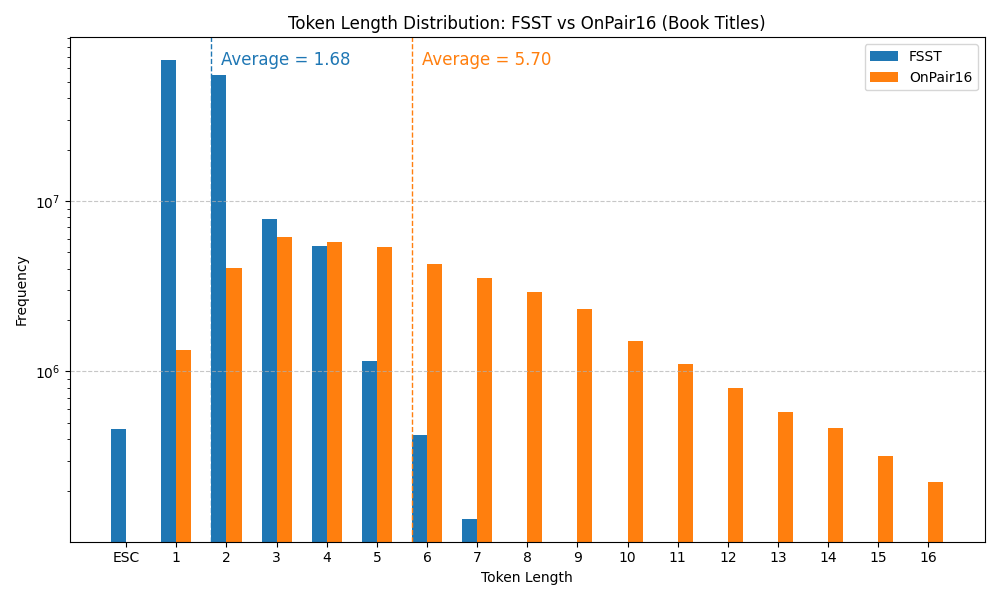}
  \caption{Token length distribution of the compressed outputs for \compressor{FSST} and \compressor{\algorithmname{16}} on the \dataset{Book Titles} dataset. \compressor{\algorithmname{16}} produces significantly longer tokens than \compressor{FSST}, reducing the number of dictionary lookups required during decompression. % Although \compressor{FSST} benefits from L1 cache dictionary storage, \compressor{\algorithmname{16}}'s longer tokens help compensate for the higher L2 cache access latency by requiring substantially fewer total dictionary accesses.
  }
  \label{fig:token_length_distribution}
\end{figure}

The average token length in the compressed output is a critical factor influencing decompression performance, as it directly determines the number of dictionary lookups required. \autoref{fig:token_length_distribution} compares the token length distribution of \compressor{FSST} and \compressor{\algorithmname{16}} for the \dataset{Book Titles} dataset. 

\compressor{\algorithmname{16}} achieves longer average token lengths ($3.36\times$ larger across all datasets) compared to \compressor{FSST}. While \compressor{FSST} benefits from storing its compact dictionary entirely in the L1 cache (access latency of 1 ns \cite{hennessy2019architecture}), \compressor{\algorithmname{16}}'s larger dictionary primarily resides in the L2 cache (access latency of 3-10 ns \cite{hennessy2019architecture}). However, the longer tokens used by \compressor{\algorithmname{16}} result in proportionally fewer dictionary accesses, which helps compensate the increased latency and allows the decompression performance to remain competitive.

\subsubsection{Token Frequency Skewness} 
\begin{figure}
  \centering
  \includegraphics[width=\linewidth]{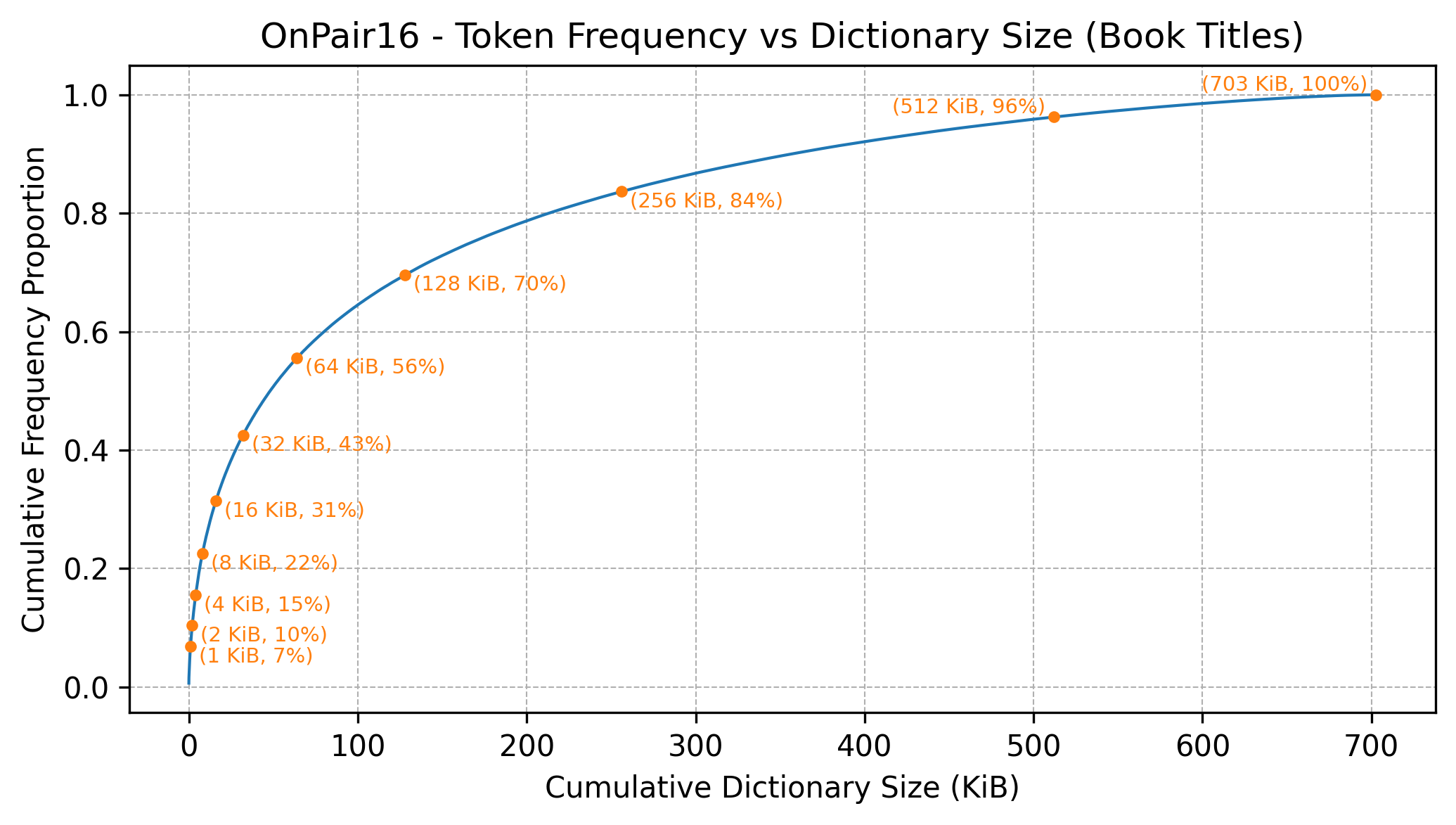}
  \caption{Cumulative token coverage vs. dictionary memory footprint for \compressor{\algorithmname{16}} on the \dataset{Book Titles} dataset. The curve shows the percentage of token occurrences (y-axis) accounted for by progressively larger subsets of the dictionary (x-axis), sorted by descending frequency. Markers highlight key thresholds; for example, 32~KiB is enough to store the most frequent tokens covering 43\% of the entire compressed output.}
  \label{fig:dictionary_skewness_onpair16}
\end{figure}

To understand the efficiency of \compressor{\algorithmname{16}}'s dictionary usage, we analyze the skewness of token frequencies in the compressed output. \autoref{fig:dictionary_skewness_onpair16} illustrates the cumulative coverage of token occurrences as progressively larger subsets of the dictionary (sorted by frequency) are included. It presents a skewed distribution where a small subset of tokens dominate the compressed output. This skew ensures that most frequently accessed tokens remain in higher levels of cache. For instance, the top 256~KiB (covering 84\% of tokens) fits comfortably within modern L2 caches (typically 256~KiB-1~MiB per core), minimizing cache misses during decompression.

\subsection{Training vs. Parsing Time Breakdown}
\autoref{tab:training_vs_parsing_time} reports the breakdown of compression time into the training and parsing phases for \compressor{\algorithmname{}} and \compressor{\algorithmname{16}} across all datasets. Parsing dominates overall compression time in almost all cases because it applies longest prefix matching to the entire dataset, while training operates on a small random sample and terminates once the dictionary reaches its target size.

The imbalance becomes more pronounced with larger datasets such as \dataset{URLs}, where dictionary construction processes only a small portion of the data, making training costs nearly negligible compared to parsing. For smaller datasets, the difference narrows since more of the input must be scanned to populate the dictionary.

\compressor{\algorithmname{16}} helps mitigate worst-case parsing costs on inputs with long, repeated prefixes—such as \dataset{URLs}. These inputs tend to produce large LPM buckets with many near-duplicate suffixes, increasing lookup time. By bounding both bucket size and dictionary entry length, \compressor{\algorithmname{16}} reduces the cost of handling these cases. This is reflected in the much lower parsing time observed on the \dataset{URLs} dataset compared to its unbounded counterpart.

In general, the parsing phase remains the dominant component of total compression time, but thanks to these optimizations, it scales well across datasets and maintains high throughput even in less favorable scenarios.

\begin{table}
  \caption{Breakdown of compression time into training and parsing phases for \compressor{\algorithmname{}} and \compressor{\algorithmname{16}} across all datasets. Parsing tends to dominate overall runtime, especially for larger files, as dictionary construction touches only a small fraction of the total data compared to parsing, whereas for smaller files the two costs are more balanced.}
  \label{tab:training_vs_parsing_time}
  \centering
  \begin{tabular}{llrr}
    \toprule
    Dataset & Compressor & Training (s) & Parsing (s) \\
    \midrule
    \rowcolor{gray!25} \dataset{Book Reviews} & & & \\
        \rowcolor{gray!0} & \compressor{\algorithmname{}}   & 0.31 & 3.44 \\
        \rowcolor{gray!5} & \compressor{\algorithmname{16}} & 0.28 & 1.89 \\
    \rowcolor{gray!25} \dataset{Book Titles} & & & \\
        \rowcolor{gray!0} & \compressor{\algorithmname{}}   & 0.29 & 1.61 \\
        \rowcolor{gray!5} & \compressor{\algorithmname{16}} & 0.28 & 0.96 \\
    \rowcolor{gray!25} \dataset{News Headlines} & & & \\
        \rowcolor{gray!0} & \compressor{\algorithmname{}}   & 0.18 & 0.30 \\
        \rowcolor{gray!5} & \compressor{\algorithmname{16}} & 0.18 & 0.18 \\
    \rowcolor{gray!25} \dataset{Tweets} & & & \\
        \rowcolor{gray!0} & \compressor{\algorithmname{}}   & 0.24 & 0.89 \\
        \rowcolor{gray!5} & \compressor{\algorithmname{16}} & 0.21 & 0.51 \\
    \rowcolor{gray!25} \dataset{URLs} & & & \\
        \rowcolor{gray!0} & \compressor{\algorithmname{}}   & 1.42 & 133.13 \\
        \rowcolor{gray!5} & \compressor{\algorithmname{16}} & 0.74 &   9.20 \\
    \bottomrule
  \end{tabular}
\end{table}

\section{Conclusions}
We introduced \compressor{\algorithmname{}} and \compressor{\algorithmname{16}}, two compression algorithms that fill the performance gap between \compressor{FSST} and \compressor{BPE} for short string datasets in main-memory database systems. Whereas \compressor{FSST} offers high speed at the expense of compression ratio, and \compressor{BPE} achieves strong compression but suffers from prohibitive training costs, \compressor{\algorithmname{}} offers a compelling trade-off: \compressor{BPE}-level compression ratios with dramatically faster training and far lower memory usage. Central to this design is a cache-efficient, single-pass dictionary construction process that operates on a small sample of the data—enabling fast deployment even on large datasets—and an optimized longest prefix matching routine ensuring parsing efficiency. The \compressor{\algorithmname{16}} variant further accelerates decompression and random access through hardware-friendly fixed-length tokens.

These properties make \compressor{\algorithmname{}} and \compressor{\algorithmname{16}} a practical choice for compress-once–decompress-many workloads in in-memory data management. By combining compact dictionaries with predictable, SIMD-friendly decoding, they help reduce memory footprint and access latency.

% \balance 

\bibliographystyle{ACM-Reference-Format}
\bibliography{sample}

\end{document}